\documentstyle[aps,epsf,twocolumn]{revtex}
\def\be{\begin{eqnarray}}
\def\ee{\end{eqnarray}}
\newcommand{\z}{\bar{z}}

\begin{document}


\title{\bf   Macroscopic Universality :\\
        \bf Why QCD in Matter is Subtle ?}

\author{ {\bf Romuald A. Janik}$^1$, {\bf Maciej A.  Nowak}$^{1,2}$ ,
{\bf G\'{a}bor Papp}$^{3}$  and {\bf Ismail Zahed}$^4$}

\address{$^1$ Department of Physics, Jagellonian University, 
30-059 Krakow, Poland.\\
$^2$ GSI, Plankstr. 1, D-64291 Darmstadt, Germany \& \\
 Institut f\"{u}r Kernphysik, TH Darmstadt, D-64289 Darmstadt, Germany \\
$^3$GSI, Plankstr. 1, D-64291 Darmstadt, Germany \& \\
Institute for Theoretical Physics, E\"{o}tv\"{o}s University,
Budapest, Hungary\\
$^4$Department of Physics, SUNY, Stony Brook, New York 11794, USA.}
\date{\today}
\maketitle

\begin{abstract}
We use a chiral random matrix model with $2N_f$ flavors to mock up
the QCD Dirac spectrum at finite chemical potential. We show that the $1/N$
approximation breaks down in the quenched state with spontaneously 
broken chiral symmetry. The breakdown condition is set by the divergence 
of a two-point function, that is shown to follow the general lore of 
macroscopic universality. In this state, the fermionic fluctuations
are not suppressed in the large $N$ limit.

\end{abstract}

{\bf 1.\,\,\,}
In the presence of a chemical potential, the lattice QCD Dirac operator is 
non-hermitean. As a result, the fermionic determinant (probability measure) 
of the QCD Dirac operator is 
a complex number. Since Monte-Carlo simulations demand a positive definite 
measure, straightforward algorithms have failed \cite{KANAYA}. Unconventional 
QCD algorithms have been devised, leading to a chiral phase transition at 
finite chemical potential and  strong coupling 
($g=\infty$) \cite{KANAYA,UNQUENCHED,GOCKSCH}. In the quenched 
approximation, however, 
the restoration of chiral symmetry was found to set in at surprisingly small
chemical potentials  of the order of half the pion mass, thereby vanishing 
in the chiral limit \cite{OTHERS}. The origin of the discrepancy was traced back to the 
phase of the Dirac operator \cite{GIBBS}, as demonstrated by in
\cite{GOCKSCH} using lattice simulations. 

While many of the basic issues related to the quenched calculations 
are known on the lattice \cite{UNQUENCHED,GOCKSCH,OTHERS,GIBBS}, we
feel the need for a better understanding using simple models.
The spontaneous breaking of chiral symmetry in QCD reflects  the 
spectral distribution of quark eigenvalues near zero-virtuality\cite{BANKS}.
For sufficiently random gauge
configurations, the Dirac operator can be approximated by a chiral 
matrix with random entries \cite{VERZA,VER}, in qualitative agreement
with a number of lattice simulations for both the bulk eigenvalue 
distributions \cite{USLATTICE}, and correlations \cite{JAC,HALASZ}.

Chiral random matrix models with finite chemical potential  have been 
discussed recently in \cite{STEPHANOV,USNJL,HALJAC} using various methods.
For zero chemical potential, the ground state breaks spontaneously chiral 
symmetry. The low-lying spectrum is that of constituent quarks with no Fermi 
surface. Naively one would expect that for a chemical potential of the order of 
the constituent quark mass, chiral symmetry would be restored. This expectation 
is not borne out by direct numerical calculations which show that chiral 
symmetry is restored for any nonzero chemical potential in the chiral limit 
and for large matrix sizes \cite{STEPHANOV}, in qualitative agreement with the
quenched lattice simulations. The numerical results have been shown to
follow from a modified replica trick \cite{STEPHANOV,SOMMERS}, in agreement
with \cite{GOCKSCH}.

In this letter, we reanalyze the spectral density with $2N_f$ conventional 
quarks. In section~2, we introduce the model and discuss the $1/N$ expansion
around the quenched state with spontaneously broken chiral symmetry. 
In section~3, we evaluate a pertinent two-point correlation in the same state,
and show that it diverges in the eigenvalue plane in the small eigenvalue
region. The support for the ensuing spectral distribution is in agreement
with a recent result \cite{STEPHANOV} using different arguments. We explain why.
Our result for the two-point correlator, extends the macroscopic 
universality argument in random matrix models \cite{BZ} to the chiral case.
In section~4 we discuss the difference between the unquenched and quenched 
spectral densities. We argue that the quenched state with
spontaneous symmetry breaking does not suppress the fermionic fluctuations.
Our conclusions are summarized in section~5.

\vskip .2cm
{\bf 2.\,\,\,}
A model form of the QCD Hamiltonian in the chiral basis with finite 
chemical potential $\mu$ and zero current quark mass is
\be
{\bf Q}_5^{QCD} \equiv {\bf M} =
\left(\matrix{ 0 &  \mu \cr
-\mu  & 0 \cr}\right) + \left(\matrix{ 0 & -i{\bf A}\cr
i{\bf A}^{\dagger} & 0 \cr}\right)
\label{KALKREUTER}
\ee
where only constant quark modes have been retained\cite{USNJL}.
In the context of four dimensional field-theories
${\bf Q}_5$ is the Hamiltonian of a Dirac particle in five dimensions.
The translation of this operator to chiral random matrix models is achieved
by assuming that the gluonic configurations ${\bf A}$ are sufficiently random 
so that ${\bf A}\rightarrow {\bf R}$, where ${\bf R}$ is an $N\times N$
complex random 
matrix with Gaussian weight. With this in mind, (\ref{KALKREUTER}) is the sum 
of a deterministic ${\bf D}$ and random (chiral) piece ${\bf R}$. 
The deterministic piece is non-hermitian for $\mu\neq 0$, with complex 
$\pm i|\mu|$ eigenvalues. 

Now, consider the  partition function associated with 
(\ref{KALKREUTER}) for $2N_f$ flavors
\be
{\bf Z} (2N_f, \mu) = <{\rm det}^{2N_f} (z-{\bf M})>
\label{ZN}
\ee
with $V_{2N_f} = -{\rm log \,} {\bf Z}/2N_f$ playing the role of a complex 
potential. For $z=im$, (\ref{ZN}) mocks up the partition function of $2N_f$
quarks of equal mass $m$. The averaging in (\ref{ZN}) is carried using a 
gaussian weight
\be
<\ldots>\equiv \frac{1}{Z} \int \ldots  e^{-N \Sigma Tr 
{\bf R}{\bf R}^{\dagger}} d{\bf R}
\label{weight}
\ee
with $\Sigma$ setting the scale of chiral symmetry breaking 
(throughout $\Sigma$ is set to 1 for simplicity, unless indicated otherwise).
The gaussian weight (\ref{weight}) is an idealization of the gluon action,
and defines the quenched averagings to be discussed below.
For sufficiently small quark eigenvalues (here constant modes), the gluon 
interaction may be thought as random \cite{VERZA}. The gaussian weight follows 
from the principle of maximum entropy.

First, let us consider the spectral distribution associated to (\ref{ZN})
in the {\it quenched} approximation. With $V_{2N_f}$ playing the role
of a complex potential, the spectral distribution follows from Gauss law
\cite{SOMMERS}
\be
\nu (z, \z ) = \lim_{N_f\to0} \rho (z, \z) = \lim_{N_f\to0}
-\frac{1}{\pi N} \partial_{\z}\partial_z V_{2N_f}
\label{QUENCHED}
\ee
Explicitly, 
$\pi\nu (z,\bar{z}) =\partial{\bf G}/\partial\bar{z}$, with the resolvent
\be
{\bf G} (z, \bar{z}) = \frac 1N \left< {\rm Tr} ({z-{\bf M}})^{-1} \right>
\label{RESOLVENT}
\ee
We have retained $(z,\bar{z})$ in both (\ref{QUENCHED}) and (\ref{RESOLVENT})
to allow for the possibility that holomorphic symmetry may be broken in 
the thermodynamical limit by the quenched averaging (\ref{weight}).
The factor of $1/N$ in (\ref{QUENCHED}) implies that $\nu (z, \z)$ is 
normalized to one on its pertinent support.

For large $z$ and to leading order in $1/N$,
${\bf G}$ obeys Pastur's equation (planar approximation) \cite{USNJL,PASTUR},
\be
{\bf G}({(z-{\bf G})^2 +\mu^2}) -{z+{\bf G}} =0
\label{PASTUR}
\ee
The solution to (\ref{PASTUR}) is a holomorphic function of $z$, except 
for line discontinuities, with end-points given by the zeroes of the 
discriminant
\be
4\mu^2 z^4 + z^2 (8\mu^4 - 20 \mu^2 -1) + 4 (\mu^2 + 1)^3 =0
\label{DISCRIMINANT}
\ee
For $\mu=0$, there  
is a cut along the real axis between $-2$ and +2. The discontinuity along the 
cut is Wigner's semicircle, with a quenched and stable 
state that breaks spontaneously chiral symmetry.
With increasing $\mu^2\rightarrow 1/8$, the two roots approach each other,
followed by the emergence of two new roots from real infinity. 
For $\mu^2=1/8$ they coalesce pairwise, and move to the complex plane 
for $\mu^2 >1/8$. This behavior suggests that the quenched state 
with $\nu(+i0, -i0)\neq 0$, supports a chiral condensate 
up to $\mu^2=0.125$, after which a first order transition is observed. This 
conclusion, however, does not hold since the $1/N$ approximation breaks down 
for small eigenvalues in the present model, as we now show.

\vskip .2cm
{\bf 3.\,\,\,}
The breakdown can be probed either by looking at the nonplanar corrections 
to (\ref{PASTUR}), or by testing the 
the stability content of the quenched state against local correlations. 
One pertinent correlation function is given by the two-point function
\be
N^2 {\bf C} (z, \z ) = <{\rm Tr}(z-{\bf M})^{-1}
               {\rm Tr}(\z-{\bf M}^{\dagger})^{-1}>_c
\label{two}
\ee
in the quenched state. The lower script in (\ref{two}) is for connected.
The physical interpretation of (\ref{two}) will be
clarified in the next section. For large values of $z$, (\ref{two})
may be analyzed using again a $1/N$ expansion \cite{BZ,WEIDEN,BZEARLY}. 
Following the arguments in \cite{BZ}, the result is \cite{USBIGPAP},
\be
N^2{\bf C} (z,\bar{z}) = -\frac{1}{4}\partial_{z}\partial_{\z}
{\rm Log} \{[({\bf H}-\mu^2)^2-|z-{\bf G}|^4]/{\bf H}^2\}
\label{logcorr}
\ee
where ${\bf H}=|z-{\bf G}|^2/|{\bf G}|^2$ and ${\bf G}$ is a 
solution to Pastur's equation (\ref{PASTUR}). For $\mu=0$ and $z=\z$
{\it outside} the cut, 
(\ref{logcorr}) reduces to $N^2 {\bf C} (z, z) = 1/{z^2 (z^2-4)^2}$,
which is different from the hermitean (nonchiral) result of
\cite{BZ}, but in agreement with the one of \cite{JAN}.
The result (\ref{logcorr}) shows that in our case the two-point correlation 
function is also amenable to the one-point function, thereby extending 
the macroscopic universality argument discussed originally in \cite{BZ} to 
the chiral case.

{}From (\ref{logcorr}) we observe that the two-point correlation function
diverges in the eigenvalue-plane in the domain prescribed by the zero of the 
logarithm,
\be
  |z-{\bf G}|^2 (1-|{\bf G}|^2)-\mu^2 |{\bf G}|^2 =0
\label{island}
\ee
For $\mu=0$, this condition is fulfilled for ${\bf G} = z$, which is a 
trivial (unphysical) solution to Pastur's equation, and for $|{\bf G}|^2=1$ a 
nontrivial (physical) solution to Pastur's equation along the cut.  The latter 
when supplemented with the condition 
$\lim_{z \rightarrow \infty}{\bf G} \sim 1/z$ in the outside,  yields 
Wigner's distribution for the eigenvalue density.

\begin{figure}[tbp]
\centerline{\epsfxsize=7.5cm \epsfbox{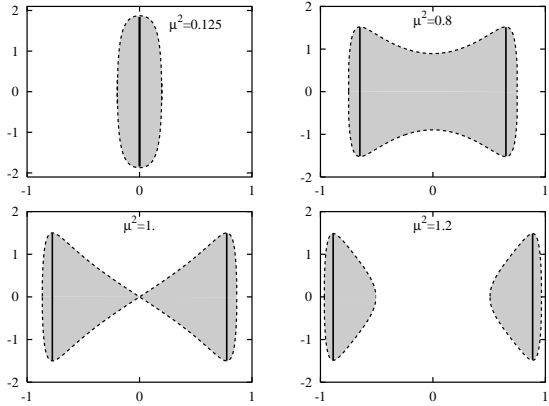}}
\caption{The envelope (dashed lines) is from 
eq.~(\protect\ref{island}) in the plane $w=iz$  while the cuts (solid lines) 
are from eq.~(\protect\ref{DISCRIMINANT}), for different $\mu$.
Shaded islands represent ``other vacuum''.}
\label{fig1}
\end{figure}

Figure~\ref{fig1} shows the envelope in dashed line for which the condition
(\ref{island}) is met in the $w=iz$ plane, for several values of the chemical 
potential $\mu$.  A structural change occurs at $\mu^2=1$.
The solid lines are the mean-field solution following 
from (\ref{DISCRIMINANT}). They lie in the eigenvalue-domain where the 
$1/N$ fluctuations are dominant, a signal that chiral symmetry is restored. 
Since (\ref{two}) mixes $z$ and $\z$, holomorphic symmetry is spontaneously
broken in the quenched state. 
One can recast~(\ref{island}) using~(\ref{PASTUR}) into ($iw=x+i y$)

\be
(\mu^2\!\!-\!\!x^2)^2 y^2=\left[ 4\mu^4(1\!\!-\!\!\mu^2)\!-\!
	(1\!\!+\!\!4\mu^2\!\!-\!\!8\mu^4)x^2\!-\!4\mu^2 x^4\right],
\label{BOUND}
\ee
The result (\ref{BOUND}) is in agreement with the one in \cite{STEPHANOV},
where the quenched spectral density was evaluated in the {\it inside}
(small eigenvalues) using a pair of conjugate quarks in the quenched 
approximation (essentially $|z-{\bf M}|$ in (\ref{ZN})). Conventional
quenched QCD in the $1/N$ approximation works as well {\it outside}.
The two approaches agree on (\ref{BOUND}) which sets their domain of 
validity. To be able to describe the spectral distribution {\it inside}
using the quenched version of (\ref{ZN}) requires a different method than 
the $1/N$ approximation we have used.

\vspace*{-2mm}

\begin{figure}[tbp]
\centerline{\epsfxsize=7.5cm \epsfbox{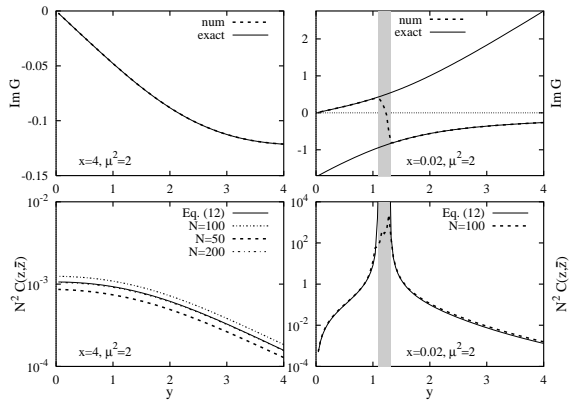}}
\caption{Imaginary part of the resolvent (upper figures) and
the correlation function (lower figures) for $w=ix-y$ with $x=4,\, 0.02$
and $y \geq 0$ at $\mu^2=2$. The solid curves are analytical
while the dashed curves are numerical (see text).}
\label{fig2}
\end{figure}

Figure~\ref{fig2} shows the analytical behavior (solid lines) for the 
imaginary part of the resolvent (\ref{RESOLVENT}-\ref{PASTUR}) 
and the correlation function (\ref{logcorr}) for two values of $w=iz$ and 
fixed $\mu^2=2$, that is $w=i4-y$ (left) and $w=i0.02-y$ 
(right) with $y\!\geq\! 0$. 
The dashed curves are a comparison to a numerical estimate using an 
ensemble of 200 chiral gaussian plus deterministic matrices with dimension 
$100\times 100$. Note that in the right figures, two different
solutions to (\ref{PASTUR}) have to be used while crossing the region of 
non-analyticity (shaded region). In the ``bounded region" between the islands 
of  Figure~\ref{fig1} the asymptotic condition
$\lim_{z \rightarrow \infty}{\bf G} \sim 1/z$ 
is no longer required. We remark that ${\rm Im}{\bf G} =0$ for $z=0$, and 
chiral symmetry is restored.

\vspace*{-2mm}

\vskip .2cm
{\bf 4.\,\,}
A dequenching of the spectral density (\ref{QUENCHED}) can be simply achieved 
by noting that since $(z-{\bf M})$ is non-hermitean, we can split it into
a phase and a modulus through
\be
(z-{\bf M})^2 = |z-{\bf M}|^2\times (\frac{z-{\bf M}}{\z-{\bf 
M}^{\dagger}})
\label{split}
\ee
In terms of (\ref{ZN}) and (\ref{split}), the {\it unquenched} spectral 
density for finite $N$ and $N_f$ is now given by $\rho (z, \z)$. From 
(\ref{ZN}) the contribution of the modulus is
\be
\rho_{mod}(z,\z)&=&\nu (z, \bar{z}) -
           \frac{N_f N}{2\pi} C(z,\bar{z}) \nonumber\\
           &+&\frac {2N_f}{N}<{\rm Tr}\delta (z-{\bf M}) \,
                    {\rm ln\,\,det} |z-{\bf M}|>_c + ...
\label{twomod}
\ee
while the contribution of the phase is
\be
\rho_{ph} (z, \z) =
           \frac{N_f N}{2\pi} C(z,\bar{z}) + ...
\label{twophase}
\ee
The density $\rho (z, \z)$ is the sum of (\ref{twomod}) and
(\ref{twophase}) modulo an extra contribution from the crossed terms,
essentially the last term in (\ref{twomod}) to the order quoted.
We note that the connected two-point function (\ref{two}) is essentially
the fluctuation of the phase of the fermion determinant in the quenched state.
It appears with opposite signs in the modulus and the phase, and cancels in the 
sum. The difference $\rho(z, \bar{z}) -\nu(z,\bar{z})$ accounts for the effects
of the {\it sea} fermions (unquenching). This difference involves nonlocal
correlation functions of $z$ and $\z$, of which (\ref{two}) is a generic 
example. The fermionic effects are down by $1/N$, provided that the correlation
functions are stable. This is not the case in the quenched state with 
spontaneous symmetry breaking.

Further insights to the above results can be achieved, if were to note that
(\ref{two}) may be rewritten as
\be 
N^2 {\bf C} (z, \z) = \langle q^{\dagger} q\,\,Q^{\dagger} Q\rangle_{\rm c}  
\label{corr}
\ee
with 
\be
qq^{\dagger} =\frac{\bf 1}{z-{\bf M}} \qquad{\rm and}\qquad
QQ^{\dagger} =\frac{\bf 1}{\z-{\bf M}^{\dagger}}
\label{PROP}
\ee
playing the role of ``propagators". 
For $\mu=0$, $\langle q^{\dagger} q\rangle$ and $\langle
Q^{\dagger}Q\rangle$ are nonzero for $z\sim i0$, and  
the vacuum contributions cancel out. Typically,
$\rho(z, \z) -\nu(z, \z)\sim {\cal O} (N_f/N)$ for $z\sim i0$.
For $\mu=0$ and large $N$, the fermionic fluctuations are suppressed 
in the quenched state. The latter breaks spontaneously chiral symmetry, 
and preserves holomorphic symmetry ($q\leftrightarrow Q$).

For small $\mu$,  (\ref{corr}) through (\ref{island}) diverges
when closing on the dashed curve of Figure~1 from the outside. This is a
signal that (\ref{corr}) 
is receiving increasingly large contributions from the ``mixed condensates" 
$\langle q^{\dagger} Q\rangle$ and their conjugates (other ``vacuum"),
so the fluctuations are not suppressed any more. 
As a result  $\rho(z, \z) -\nu(z, \z)\sim 
{\cal O} (N_f N^0)$ for small $z$. In the quenched state with spontaneous 
symmetry breaking, the fermionic fluctuations are not suppressed in the 
large $N$ limit.

\vskip .2cm
{\bf 5. \,\,} 
Using a chiral random matrix model with a finite chemical potential $\mu$, we 
have shown that the small and quenched eigenvalue distribution of the Dirac 
operator is fluctuation driven, with a size in the complex plane conditioned 
by the divergence of the connected two-point correlation function (\ref{two}). 
In this domain the $1/N$ approximation breaks down and the effects from the 
fermionic fluctuations do not decouple in the thermodynamical limit. The 
fluctuation driven phase breaks spontaneously holomorphic symmetry and 
preserves chiral symmetry, as originally shown in \cite{STEPHANOV}.

\vglue 0.2cm
{\bf \noindent  Acknowledgments \hfil}
\vskip 0.1cm
We thank Jurek Jurkiewicz and Jac Verbaarschot for discussions.
M.A.N. thanks the INT at the University
of Washington for its hospitality during the completion of this work.
This work was supported in part  by the US DOE grant DE-FG-88ER40388,
 by the Polish Government Project (KBN)  grants  2P03B19609 and
2P03B08308 and by the Hungarian Research Foundation OTKA. 

\setlength{\baselineskip}{15pt}

\end{document}